# Physics-informed operator learning for transferable energy-dissipative microstructure dynamics

Jie Xiong[a, b, 1*], Yue Wu[a,1], Xuewei Zhou[d], Peishuo Zhao[d], Jiaming Zhu[c,d,e*]

[a] Materials Genome Institute, Shanghai University, Shanghai 200444 China

[b] State Key Laboratory of Materials for Advanced Nuclear Energy, Shanghai University, Shanghai 20044, PR China

[c] State Key Laboratory of Nonlinear Mechanics, Institute of Mechanics, Chinese Academy of Sciences, Beijing 100190, China.

[d] School of Civil Engineering, Shandong University, Jinan 250061, China

[e] Shenzhen Research Institute of Shandong University, Shenzhen 518057, Guangdong Province, China

[1] These authors contributed equally to this work and should be considered co-first authors.

E-mail: xiongjie@shu.edu.cn (J. Xiong), zhujiaming@sdu.edu.cn (J. Zhu)

**Abstract**: Phase-field simulations provide mechanistic descriptions of microstructure evolution, but repeated high-fidelity integration over long horizons and broad parameter spaces remains computationally expensive. We present PFNet, a physics-informed neural operator framework that advances microstructural states by learning conditional evolution operators rather than direct correlations. PFNet combines a diffusion-inspired U-Net with periodic padding, entropy-based state conditioning and thermodynamic-parameter modulation to encode boundary consistency, instantaneous ordering state and changes in the free-energy landscape. For Cahn-Hilliard coarsening, PFNet achieves accurate one-step prediction and stable autoregressive rollouts across composition, gradient-energy coefficient, coarsening stage and morphology class, with errors concentrated near diffuse interfaces and topology-changing regions. The same framework extends to a four-channel martensitic-transformation benchmark without martensite-specific redesign. These results indicate that physics-informed operator learning can provide transferable surrogates for phase-field dynamics and broader energy-dissipative dynamical systems.

**Keyword**: Phase-field modeling; Physics-informed machine learning; Neural operator; Microstructure evolution; Diffusion-inspired neural network

## Introduction

Phase-field modeling has become a widely used theoretical and computational framework for predicting microstructure evolution in materials. It is well established that macroscopic performance of materials depend not only on chemical composition but also on microstructure, e.g. the spatial arrangement of phases, precipitates, defects, and grain boundaries[1-10]. Microstructure evolutions are governed by thermodynamic driving forces and kinetic constraints, and are often influenced by coupled fields such as composition, temperature, stress, as well as electrostatic and magnetic interactions[11-13]. The phase-field method describes microstructures using continuous order parameters, which removes the need for explicit front tracking [7, 14-19]. Within this variational setting, interface motion and morphological transitions arise from the tendency of the system to reduce an underlying free energy functional under appropriate kinetic laws[13, 14, 20]. Couplings to concentration, temperature, stress, and long-range interactions can be incorporated within the same formulation, enabling quantitative links between field conditions, transient morphologies, and structure-sensitive descriptors that underpin performance across a broad range of conditions [19, 20].

Despite its generality, phase-field simulation remains expensive for problems that demand long horizons, fine interfacial resolution, or broad parameter exploration. However, this capability is accompanied by substantial computational expense. Accurate resolution of diffuse interfaces requires fine spatial discretization, while stable integration of stiff interfacial kinetics typically demands small time steps, particularly in three-dimensional systems and in long-time coarsening where relevant dynamics unfold over many decades in time[21-25]. From a variational viewpoint, many phase-field models can be expressed as energy dissipating flows, and long-time integration repeatedly evaluates a force-like variational derivative and applies a corresponding evolution operator on the microstructure state space. This repeated state update is the dominant computational bottleneck, and it becomes particularly restrictive when one seeks systematic sweeps across composition, free energy landscapes, mobility parameters, elastic constants, or coupling strengths that govern multi-physics response.

Extensive numerical strategies have been developed to reduce this cost. Domain decomposition distributes the workload across parallel computing architectures[25]. Adaptive mesh refinement concentrates resolution near interfaces while keeping bulk regions coarse, and implicit or semi-implicit schemes alleviate restrictive stability constraints to permit larger time increments[21, 26, 27]. Fourier spectral methods exploit periodicity to compute spatial derivatives efficiently with high accuracy, and GPU acceleration further increases throughput by mapping stencil operations and spectral transforms onto massively parallel hardware[28-32]. These advances substantially improve simulation throughput, but they do not eliminate the need for repeated high-fidelity PDE integration. Routine three-dimensional studies and broad parameter-space exploration therefore remain time consuming, motivating surrogate approaches that reduce computational cost while retaining physical reliability.

Machine learning has emerged as a promising route for accelerating phase-field simulations at reduced marginal cost. Recurrent neural networks and related sequence models have reproduced representative pattern-formation scenarios including spinodal decomposition, grain growth, and dendritic solidification with high accuracy and substantial speedups [33-35]. Their practical value is often limited by challenges in long-horizon accuracy and in generalizing across conditions that shift length scales, kinetics, or phase equilibria. Neural operator approaches including Fourier Neural Operator (FNO) [36-38]. Deep Operator Network (DeepONet) [39-43], and Physics-Informed Neural Operator (PINO) [44, 45] aim to approximate the solution operator of parametric partial differential equations (PDEs) in a manner that transfers across discretization. FNO parameterizes integral kernels in Fourier space to learn global convolutions and is designed to generalize across resolutions, which aligns naturally with periodic boundary conditions (PBCs) that are common in phase-field simulations and with the need to capture long-range spatial correlations [36-38]. DeepONet provides a complementary function-to-function representation through coupled branch and trunk networks, with strong approximation capability and potential sample efficiency when high-fidelity simulation data are expensive [39-43]. Yet neither purely data-driven FNO nor DeepONet inherently guarantees physical structure such as conservation laws, monotone free-energy decay, or parameter-consistent thermodynamic responses, so their predictions often require additional verification or constraints. PINO mitigates some of these concerns by embedding governing equations or dissipation laws into the training process, but training can still be expensive. The balance between data matching and physics regularization becomes delicate in stiff, multiscale, or strongly parameter-dependent regimes. Across these paradigms, a recurring challenge is to achieve long-horizon reliability while maintaining transfer across initial conditions and parameters without incurring repeated retraining costs.

The present work develops PFNet as a physics-informed neural operator framework for thermodynamically constrained microstructure dynamics. PFNet is designed to address a central challenge in physics-informed machine learning that how to construct data-driven predictors that remain reliable during autoregressive rollout and under changes in physical parameters. Rather than learning short-time microstructure-to-microstructure correlations, PFNet learns conditional evolution operators on periodic microstructural state spaces. The framework combines a variational phase-field formulation with a diffusion-inspired U-Net, periodic padding, entropy-based state conditioning and thermodynamic-parameter modulation. These design choices encode boundary consistency, instantaneous ordering state and changes in the free-energy landscape directly into the learned update rule. We demonstrate PFNet on conserved Cahn-Hilliard coarsening and non-conserved multichannel martensitic transformation, showing transfer across composition, gradient-energy coefficient, coarsening stage, morphology class, and kinetic type. These results position PFNet as a transferable surrogate for phase-field dynamics and as a representative model for physics-embedded operator learning in energy-dissipative systems.

## Results

### PFNet learns physics-informed evolution operators

Phase-field evolution can be formulated within a unified variational framework. Let $\phi$ $(\mathbf{x}, t)$ denote a phase-field variable defined on a periodic domain $\Omega = [0, L]^2$, where $\mathbf{x} \in \Omega$ is the spatial coordinate and $t$ is the time. The physical setting determines whether $\phi$ represents a non-conserved structural order parameter or a conserved composition field. The Ginzburg-Landau free-energy functional is written as

$$\mathcal{F}[\phi] = \int_{\Omega} \left( f(\phi) + \frac{\kappa}{2} |\nabla\phi|^2 \right) d\mathbf{x} \tag{1}$$

where $f(\phi)$ is the local free-energy density and $\kappa > 0$ is the gradient-energy coefficient that penalizes interfacial gradients. The corresponding variational derivative is

$$\mu = \frac{\delta \mathcal{F}}{\delta \phi} = \frac{\partial f}{\partial \phi} - \kappa \nabla^2 \phi \tag{2}$$

where $\mu$ denotes the thermodynamic driving force, such as the chemical potential in conserved dynamics.

Allen-Cahn (AC) kinetics converts this variational force into non-conserved relaxation,

$$\frac{\partial \phi}{\partial t} = -M_{AC}\mu = -M_{AC}\left( \frac{\partial f}{\partial \phi} - \kappa \nabla^2 \phi \right) \tag{3}$$

where $M_{AC}$ is a positive kinetic coefficient. This dynamics corresponds to the $\boldsymbol{L}^2$-gradient flow of the free energy and does not conserve the spatial integral $\int_{\Omega} \phi d\mathbf{x}$. Cahn-Hilliard (CH) kinetics converts the same variational force into conserved diffusion,

$$\frac{\partial \phi}{\partial t} = \nabla \cdot [M_{CH} \nabla \mu] = \nabla \cdot \left[ M_{CH} \nabla \left( \frac{\partial f}{\partial \phi} - \kappa \nabla^2 \phi \right) \right] \tag{4}$$

where $M_{CH}$ is the mobility. Constant mobility reduces this equation to

$$\frac{\partial \phi}{\partial t} = M_{CH} \nabla^2 \left( \frac{\partial f}{\partial \phi} - \kappa \nabla^2 \phi \right) \tag{5}$$

The divergence structure guarantees the conservation of the total mass $\int_{\Omega} \phi d\mathbf{x}$.

Therefore, AC and CH kinetics differ in their conservation laws and metric structures, yet share the same free-energy functional and variational driving force. Both can be expressed in the unified operator form

$$\frac{\partial \phi}{\partial t} = -\mathcal{M}\frac{\delta \mathcal{F}}{\delta \phi} \tag{6}$$

where $\mathcal{M}$ is a positive semidefinite Onsager operator. AC kinetics uses $\mathcal{M} = M_{AC}I$ with $I$ denoting the identity operator, whereas CH kinetics employs $\mathcal{M} = -\nabla \cdot M_{CH}\nabla$. Periodic boundary conditions yield

$$\frac{d\mathcal{F}}{dt} = -\langle\frac{\delta \mathcal{F}}{\delta \phi}, \mathcal{M}\frac{\delta \mathcal{F}}{\delta \phi}\rangle \leq 0 \tag{7}$$

where $\langle\cdot,\cdot\rangle$ denotes the appropriate inner product. This inequality makes explicit the dissipative gradient-flow structure shared by AC and CH dynamics.

The operator-level formulation also admits a probabilistic interpretation. Let $p_t(\phi)$ denote the probability density of microstructural states at time $t$ induced by an ensemble of random initial conditions $\phi_0$. Deterministic phase-field evolution drives $p_t(\phi)$ through the continuity equation

$$\partial_t p_t(\phi) + \nabla_\phi \cdot \left(p_t(\phi)\mathbf{v}_t(\phi)\right) = 0 \tag{8}$$

where $\nabla_\phi$ is the functional gradient and $\mathbf{v}_t(\phi)$ is the velocity field in state space. The unified phase-field kinetics gives $\mathbf{v}_t(\phi) = -\mathcal{M}\frac{\delta \mathcal{F}}{\delta \phi}$. This representation shows that phase-field evolution moves probability mass deterministically along the manifold of microstructures.

The Boltzmann distribution $p_{eq}(\phi) \propto \exp(-\frac{\mathcal{F}[\phi]}{k_B T})$ satisfies

$$\nabla_\phi \log p_{eq}(\phi) = -\frac{1}{k_B T}\frac{\delta \mathcal{F}}{\delta \phi} \tag{9}$$

which reveals that the AC drift is rigorously proportional to the score of the equilibrium distribution

$$\mathbf{v}_t(\phi) = M_{AC}k_B T\nabla_\phi \log p_{eq}(\phi) \tag{10}$$

CH kinetics admits an analogous correspondence in the Wasserstein metric, with conservation enforced by the mobility operator. This shared variational and probabilistic structure motivates a unified surrogate design for both non-conserved and conserved phase-field systems.

Guided by the operator formulation above, we develop PFNet as a physics-embedded surrogate model for microstructural evolution. PFNet is not intended as a generic microstructure-to-microstructure mapping. Instead, it learns the phase-field evolution operator over a fixed physical interval $\Delta t$. Let the discretized microstructure at step *t* be $\phi^t \in \mathbb{R}^{C \times N \times N}$, where *C* is the number of field channels and *N* is the spatial grid size. The learned dynamical map is parameterized as

$$\phi^{t+1} = G_\theta(\phi^t, \mathcal{H}(\phi^t), \kappa) \tag{11}$$

where $\mathcal{H}(\phi^t)$ is the Shannon entropy of the current state and $G_\theta$ is a neural network with trainable parameters $\theta$.

The backbone of $G_\theta$ is a diffusion-style U-Net, chosen for structural analogy between score-based generative model and phase-field evolution. Both frameworks update high-dimensional states using a score-like vector field. Score-based models define the probability-flow ordinary differential equation as

$$d\mathbf{x} = \left[\mathbf{b}(\mathbf{x}, \tau) - \frac{1}{2} g(\tau)^2 \nabla_{\mathbf{x}} \log p_\tau(\mathbf{x})\right] d\tau \tag{12}$$

where $\tau$ is the diffusion timestep, $\mathbf{b}$ is the drift term, g is the diffusion coefficient, and $\nabla_{\mathbf{x}} \log p_\tau$ is the score. Diffusion models use this score field to move samples toward the data manifold, directly mirroring the way phase-field dynamics uses the variational derivative to drive microstructures toward lower-free-energy states.

PFNet applied circular padding throughout the network to enforce periodic boundary conditions native to representative volume elements. This representation-level constraint effectively prevents artificial boundary artifacts during long autoregressive rollouts. The present work further introduces $\mathcal{H}(\phi^t)$ as an adaptive conditioning variable,

$$\mathcal{H}(\phi^t) = -\sum_{k=1}^{K} p_k \log_2 p_k, \quad p_k = \frac{|\{i: \phi_i^t \in B_k\}|}{N^2} \tag{13}$$

where *K* = 256 is the number of discrete bins used to partition the field range, and $p_k$ is the empirical occupancy probability of the *k*-th bin $B_k$. As ordering and coarsening proceed, $\mathcal{H}(\phi^t)$ decreases and provides a compact scalar descriptor of instantaneous disorder level. Diffusion models commonly rely on an external timestep to index disorder, whereas this entropy signal is computed directly from the current microstructure. It therefore provides a trajectory-adaptive and thermodynamically meaningful conditioning signal.

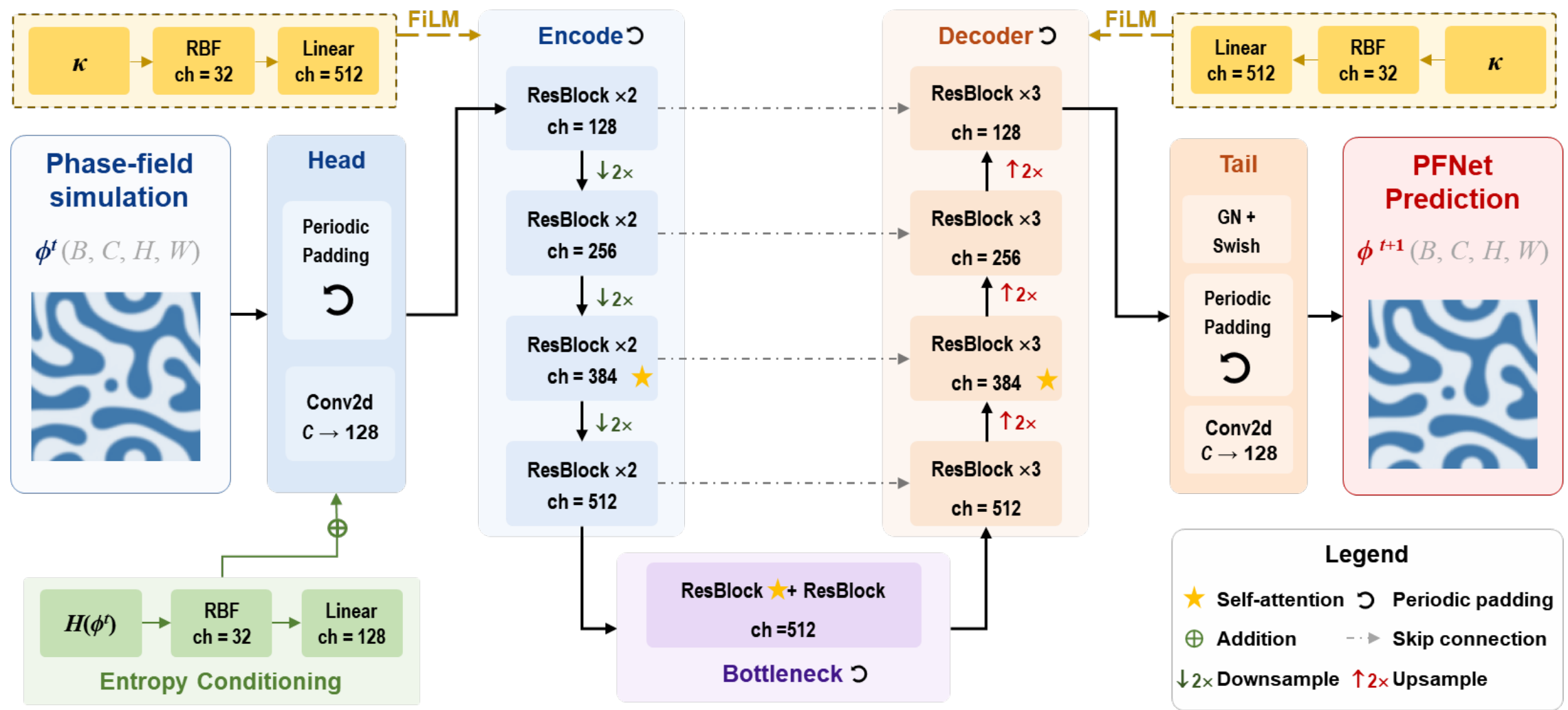

**Fig. 1** | Physics-embedded architecture of PFNet for microstructure evolution prediction. PFNet employs a diffusion-style U-Net backbone with dual physics-informed conditioning. The input phase-field state $\phi^t$ is processed through periodic-padded convolutions across four hierarchical levels (128, 256, 384, 512 channels). Entropy $H(\phi^t)$ computed from the current microstructure, is embedded via RBF and added to head features, providing trajectory-adaptive conditioning. The gradient-energy coefficient $\kappa$ is embedded and injected through feature-wise linear modulation (FiLM) in all residual blocks, enabling parameter-aware predictions. Self-attention (denoted by stars) is applied only at middle resolution (384 channels) and the first bottleneck block to capture long-range patterns efficiently. Skip connections link encoder and decoder at matching resolutions. Periodic padding enforces physical boundary conditions throughout. The architecture unifies phase-field tasks through adaptive input channels and $\kappa$ conditioning, producing one-step predictions $\phi^{t+1}$ that preserve microstructural physics.

The gradient-energy coefficient $\kappa$ also enters the network through feature-wise modulation. Since $\kappa$ directly controls the interfacial penalty in governing phase-field equation, this conditioning mechanism allows the surrogate to adapt its internal dynamics continuously across different free-energy landscapes. PFNet therefore combines multi-scale periodic convolutions, skip connections, and a bottleneck attention block to capture both local interfacial morphology and long-range coarsening patterns. More details of PFNet please see the **Supplementary Materials**.

## Single-step prediction for conserved Cahn-Hilliard kinetics

The CH equation is adopted as the benchmark system to generate high-fidelity data. Its conserved dynamics enables systematic evaluation of PFNet under a strict mass-conservation constraint. Single-step prediction results for CH kinetics are summarized in **Fig. 2**. The predicted fields agree closely with the reference solutions across the early, intermediate, and late stages of coarsening. Early dynamics at $t$ = 50 exhibits weak compositional fluctuations associated with the onset of spinodal decomposition, which PFNet recovers with accurate amplitude and spatial texture. Subsequent temporal evolution up to $t$ = 200 yields distinct phase-separated domains where the model retains the number density, spatial arrangement, and local elongation of minority regions. Longer evolution at $t$ = 500 and 800 further demonstrate that PFNet captures enlarged droplet morphologies, smooth diffuse interfaces, and the global spatial distribution of coarsened domains without introducing unphysical morphological distortions.

The training dynamics also suggest that entropy-based conditioning facilitates optimization. The training error reaches approximately $10^{-5}$ within the first 20 epochs. Ablation tests without entropy conditioning show that the model remains trainable, but requires more epochs to achieve a comparable error level. These results suggest that $\mathcal{H}(\phi^t)$ acts as an informative state descriptor, helping the PFNet distinguish different ordering stages and accelerating convergence.

Pointwise error maps further clarify this predictive fidelity. Residuals are mainly localized near diffuse interfaces and high-curvature regions, while the thermodynamically stable bulk phases remain largely unaffected. Quantitative errors in **Fig. 2b** support this observation. The single-step prediction error remains on the order of $10^{-5}$ over nearly the entire evolution period for all tested $\kappa$ values. A systematic thermodynamic parameter dependence nevertheless emerges. The $\kappa$ = 5 case shows both the largest mean error and the widest uncertainty band, whereas $\kappa \geq 7$ yields lower and more tightly clustered error distributions. This trend is consistent with the role of $\kappa$ in phase-field formulations. A smaller $\kappa$ narrows the diffuse interface width and steepens local concentration profiles. The resulting sharper interfacial gradients amplify localized chemical potential variations and make the evolution more challenging to resolve numerically (see **Supplementary Materials**).

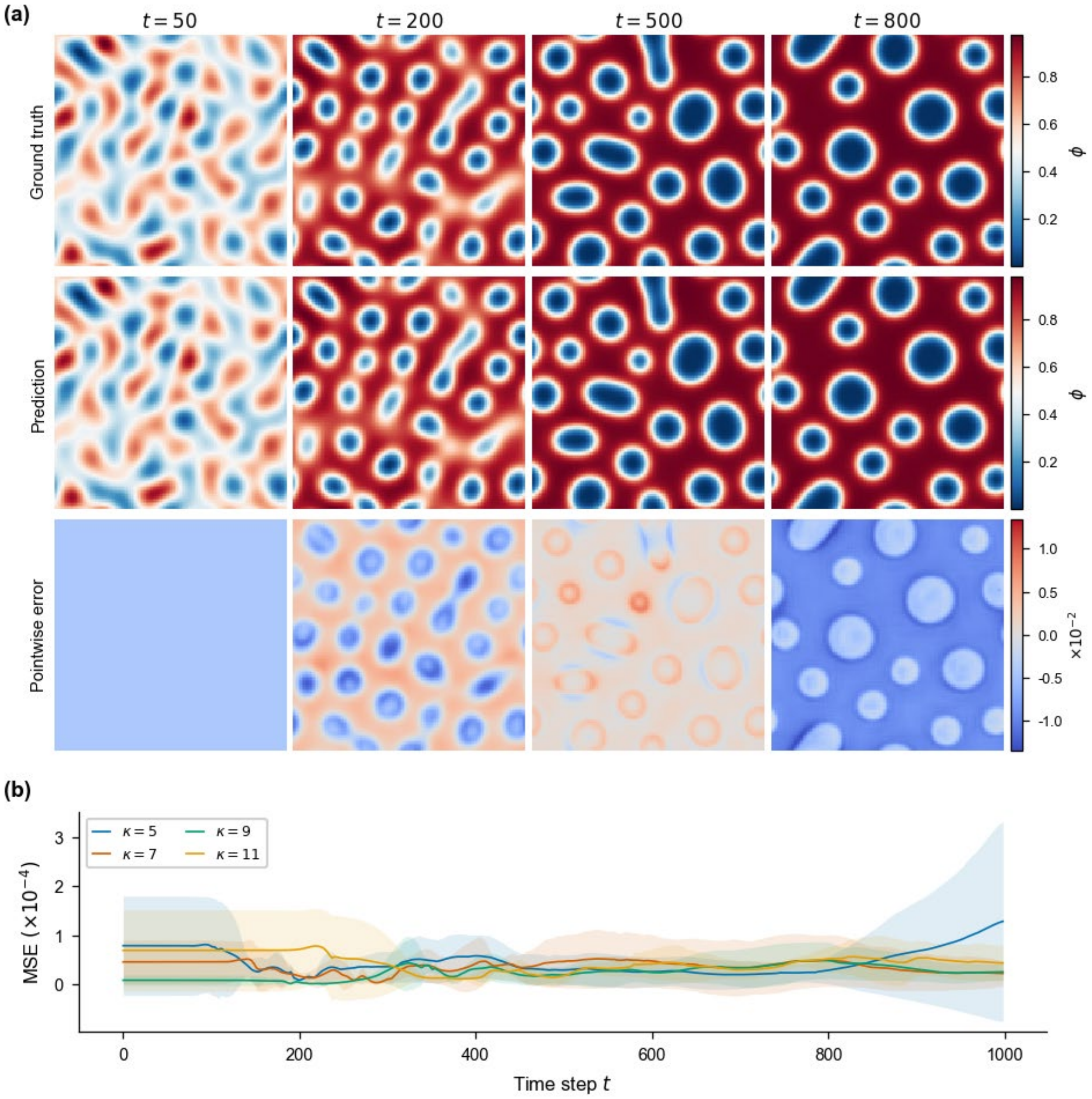

**Fig. 2** | Single-step prediction for Cahn-Hilliard (CH) kinetics. **(a)** Ground-truth concentration fields, PFNet predictions, and pointwise errors at $t$ =50, 200, 500, and 800, covering the early, intermediate, and late stages of coarsening. The predicted fields remain close to the reference solutions throughout the evolution, with errors mainly localized near diffuse interfaces and high-curvature regions. (b) Single-step test error as a function of time step for different gradient-energy coefficients $\kappa = 5$, 7, 9, and 11. Solid lines denote the mean error and shaded bands denote the spread over the test set.

## Long-horizon rollout and morphology-level generalization

Before examining PFNet rollouts, we evaluated a DeepONet baseline. Despite achieving reasonable one-step accuracy, DeepONet rapidly accumulates autoregressive error and reaches order-unity deviation within a few tens of updates. This early divergence indicates a departure from the physically admissible microstructural manifold. The comparison shows that low one-step error alone is insufficient for stable long-horizon phase-field prediction. PFNet was therefore evaluated through autoregressive rollouts over progressively longer horizons after its one-step accuracy had been established.

**Fig. 3a** compares predicted and reference microstructures initialized at $t = 250$ and advanced over rollout horizons up to $\Delta t = 250$. The structural agreement remains robust up to $\Delta t = 100$. PFNet reproduces the dissolution of small thermodynamically unstable droplets, the Ostwald ripening of larger domains, and the progressive smoothing of diffuse interfaces driven by capillary forces. The corresponding pointwise errors remain weak and are largely confined to interfacial zones. The characteristic domain length scale, phase connectivity, and global coarsening pathway are preserved within this horizon.

Discrepancies become more visible as the rollout horizon increases, yet their spatial distribution remains physically interpretable. The first major deviations appear near shrinking islands, thin ligaments, and incipient coalescence sites at $\Delta t = 150$, 200 and 250. These locations coincide with regions of high local curvature and steep chemical-potential gradients. Small positional shifts of diffuse interfaces therefore produce disproportionately large pointwise residuals. The resulting degradation remains local rather than global. PFNet still preserves the dominant morphology class and the overall topological coarsening trajectory at $\Delta t = 250$. The primary limitation is therefore fine-scale interfacial registration rather than a collapse of the macroscopic microstructural topology.

Similar stability is observed when the initial composition is varied at a fixed rollout horizon of $\Delta t = 50$. **Fig. 3b** includes droplet-dominated structures, bicontinuous labyrinthine patterns, and inverted droplet morphologies across different nominal Nb fraction. PFNet reproduces the primary microstructural features across these distinct topological regimes. Error residuals remain concentrated near interfaces and do not propagate onto the stable bulk phases, demonstrating that the learned nonlinear evolution operator generalizes beyond a single predefined morphology class.

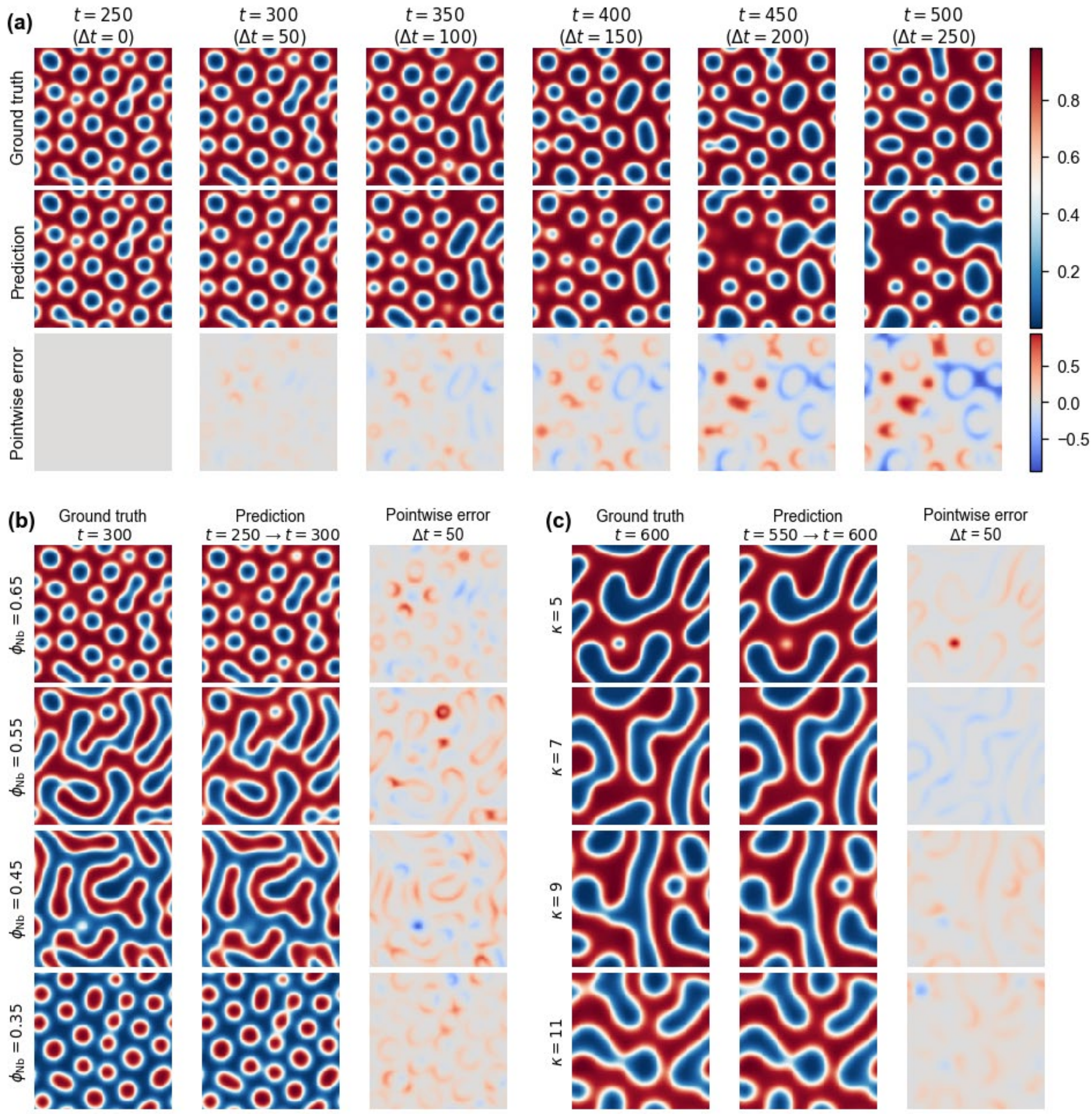

**Fig. 3**. | Representative long-horizon rollouts for CH kinetics. (a) Ground-truth fields, autoregressive PFNet predictions initialized at $t = 250$, and pointwise errors at rollout horizons $\Delta t$ = 1, 50, 100, 150, 200, and 250. PFNet preserves the dominant coarsening pathway over extended horizons, while the first visible deviations appear near shrinking droplets, thin ligaments, and incipient coalescence sites. (b) Generalization across different nominal compositions, $\phi_{\mathrm{Nb}} = 0.35$, 0.45, 0.55, and 0.65, at a fixed rollout horizon of $\Delta t = 50$. Shown are the ground truth at $t = 300$, the prediction from $t = 250$ to 300, and the corresponding pointwise error. (c) Generalization across different gradient-energy coefficients, $\kappa$ = 5, 7, 9, and 11, at a fixed rollout horizon of $\Delta t = 50$. Shown are the ground truth at $t = 600$, the prediction from $t =$ 550 to 600, and the corresponding pointwise error.

**Fig. 3c** extends this analysis to variations in $\kappa$ at the same rollout horizon of $\Delta t =$ 50. PFNet captures broad and smooth compositional channels associated with thicker interfaces, as well as finer droplet-like features associated with sharper interfaces. The predicted interface positions maintain alignment with the reference solutions, keeping the error maps structurally weak and spatially sparse. Since changing $\kappa$ alters the underlying Gibbs free-energy functional rather than merely shifting the initial condition, this result suggest that PFNet has internalized a parameter-conditioned phase-field evolution rule.

**Statistical evaluation of long-horizon rollout**

We further quantify the autoregressive behavior over a fixed 100-step horizon. The statistical distributions in **Fig. 4a** confirm the robustness observed in the representative rollouts. PFNet maintains bounded rollout errors with a clear dependence on $\kappa$ across all tested systems. Errors remain clustered and plateau near $10^{-3}$ for $\kappa \geq 7$, indicating stable long-horizon evolution. The $\kappa = 5$ case shows a higher mean error and a broader uncertainty distribution, spanning approximately $10^{-4}$ to $10^{-1}$. This trend is consistent with the governing thermodynamics. A smaller $\kappa$ enforces narrower diffuse interfaces and steeper local chemical potential gradients, which increase the intrinsic difficulty of long-horizon phase-field prediction.

The broadened error distribution for $\kappa = 5$ further suggests that rollout difficulty is controlled by more than interfacial width alone. Variations in the initial average composition generate distinct topological regimes that require different degrees of nonlinear microstructural rearrangement, as shown in **Supplementary Materials**. Some trajectories remain well controlled over 100 steps, whereas others accumulate substantial interfacial drift whenever the microstructure undergoes sensitive necking or coalescence events. The observed statistical dependence on both $\kappa$ and nominal composition is therefore consistent with the physical complexity of CH coarsening.

The 100-step test represents a demanding evaluation window for phase-field surrogate modeling. The DeepONet baseline accumulates autoregressive error rapidly and diverges much earlier, indicating that the predicted states leave the physically admissible microstructural manifold. PFNet avoids this structural failure, remains stable over the full 100-step horizon, and continuously produces morphologically meaningful phase states.

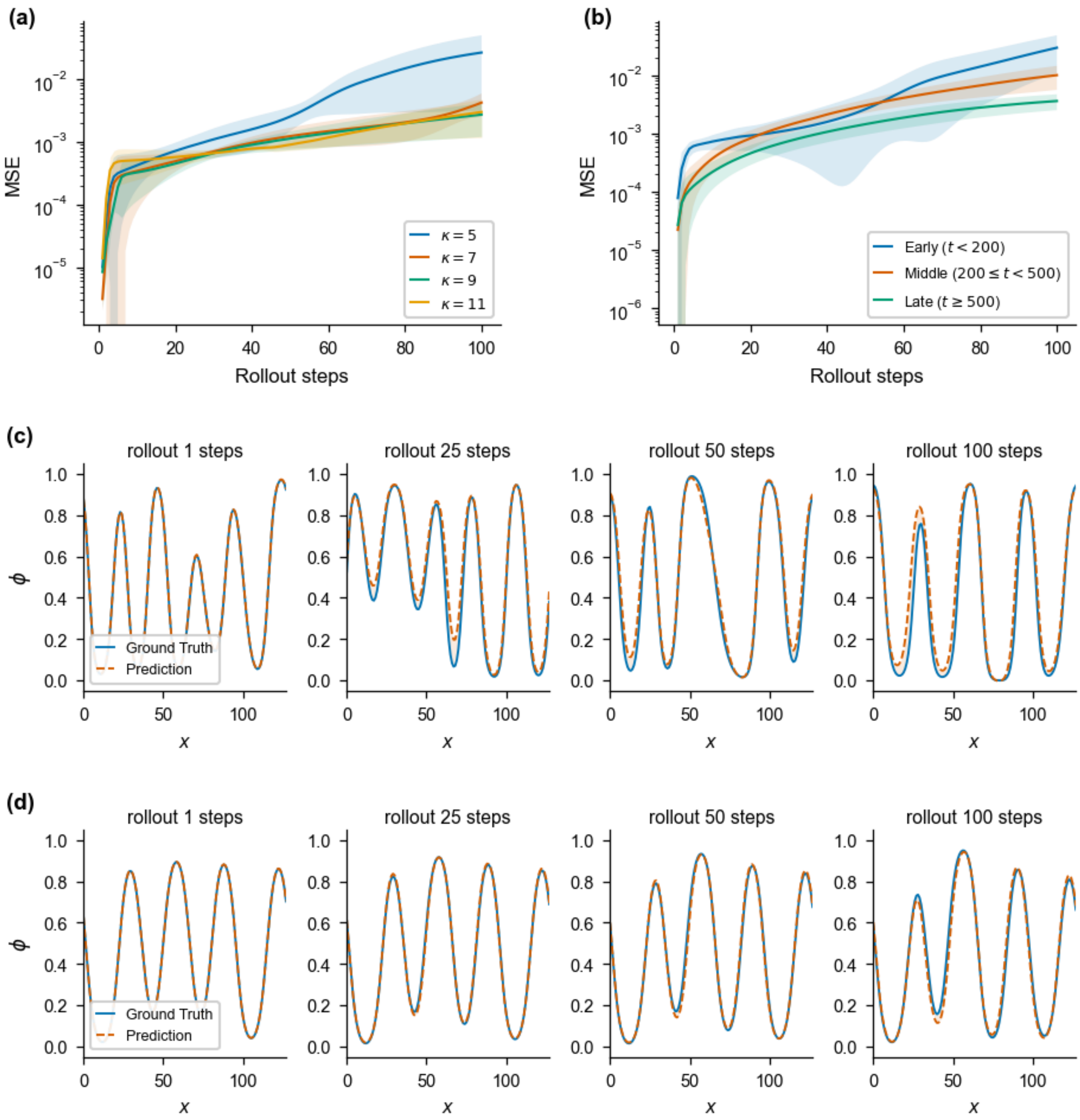

**Fig. 4.** | Quantitative evaluation of long-horizon rollout for CH kinetics. **(a)** Rollout error as a function of rollout steps for different gradient-energy coefficients, $\kappa$ = 5, 7, 9, and 11. Solid lines denote the mean error and shaded bands denote the spread over the test set. **(b)** Rollout error grouped by the starting stage of the trajectory, with early ($t < 200$), middle ($200 \leq t < 500$), and late ($t \geq 500$) initial states. Early-stage configurations show the fastest error accumulation, whereas late-stage configurations remain the most stable. **(c)** Comparison of concentration profiles ($\phi_{\mathrm{Nb}}$ = 0.55) extracted across the steepest interface for the sharper-interface case $\kappa$ = 5 after 1, 25, 50, and 100 rollout steps. **(d)** Corresponding concentration-profile comparison for the smoother-interface case $\kappa$ = 11.

Rollout stability also depends on the evolutionary stage of the phase-field trajectory. Initial states extracted from the earliest stages of spinodal decomposition, where $t < 200$, exhibit the fastest error accumulation. Intermediate states show moderate error growth, whereas late-stage structures remain the most stable. This hierarchy follows the intrinsic non-equilibrium dynamics of the CH equation. Early evolution is governed by active uphill diffusion and strong topological sensitivity, while late-stage evolution is dominated by slower curvature-driven Ostwald ripening. The microstructural trajectory becomes more predictable as the global thermodynamic driving force dissipates, allowing the learned evolution operator to maintain greater autoregressive stability.

Profile-level concentration comparisons further confirm that PFNet preserves the local interfacial structure over long autoregressive rollouts. For the sharper-interface case with $\kappa = 5$, shown in **Fig. 4c**, the predicted concentration profiles remain almost indistinguishable from the reference profiles over 1, 25, and 50 rollout steps. Even after 100 steps, the main peaks and valleys are retained, with only a slight local shift near selected interfaces. The smoother $\kappa = 11$ case in **Fig. 4d** shows similarly strong agreement, with the predicted and reference profiles nearly overlapping throughout the full 100-step horizon. These results indicate that the rollout error observed is primarily associated with small localized interface-registration errors rather than a global collapse of the concentration field. PFNet therefore maintains the dominant concentration modulation and interfacial morphology over a demanding long-horizon rollout, although small end-stage composition drift should still be assessed separately through the spatially averaged Nb concentration.

### Extension to multichannel martensitic transformation

Martensitic transformation differs fundamentally from CH coarsening in both order-parameter structure and kinetic character. CH dynamics describes conserved compositional evolution driven by chemical-potential gradients, whereas martensitic transformation is typically governed by AC-type kinetic for non-conserved structural order parameters coupled through elastic and crystallographic constraints. This contrast provides a more stringent test of whether PFNet can extend form conserved scalar diffusion to non-conserved, multichannel structural evolution under the unified variational framework established above.

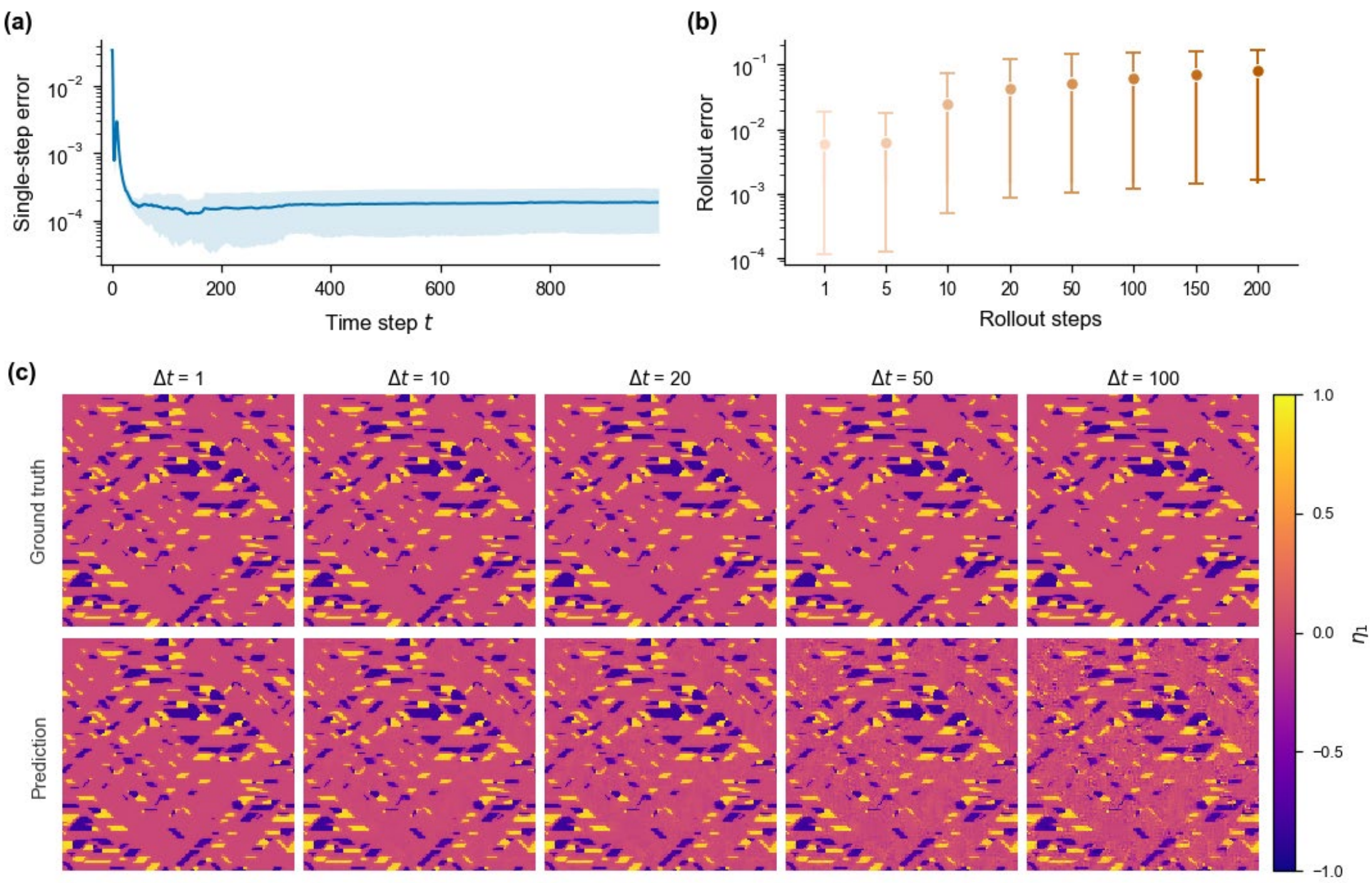

**Fig. 5**. Extension of PFNet to multichannel martensitic transformation. **(a)** Single-step error as a function of time step for the martensitic benchmark. After the initial transient, the error decreases rapidly and remains at a low level. **(b)** Rollout error as a function of rollout steps for the martensitic benchmark. Markers denote mean errors and error bars denote the spread over the test set with different start time. **(c)** Ground-truth and predicted fields for a representative order parameter $\eta_1$, at rollout horizons $\Delta t$ = 1, 10, 20, 50, and 100.

Here, we further examine PFNet using a four-variable ($\eta_1$ to $\eta_4$) phase-field model of martensitic transformation. The microstructural state is represented by four coupled non-conserved order-parameter channels associated with multivariant martensitic domains. Further details of the martensitic phase-field model and simulation protocol can be found in [46-48]. The same backbone, entropy-based conditioning, circular padding, and autoregressive rollout strategy are retained. The only structural change is that the input-output tensor is expanded from B × 1 × H × W to B × 4 × H × W. The $\kappa$-injection branch is removed because variation in $\kappa$ is not considered in this benchmark. The martensitic case therefore tests whether PFNet learns a general microstructural evolution rule rather than a scalar-specific predictor for CH coarsening.

The single-step error in Fig. 5a is elevated only at the earliest stage of the evolution, after which it decreases rapidly and stabilizes around $10^{-4}$. Early martensitic transformation involves rapid variant selection and local twin nucleation, whereas later evolution is dominated more by migration and refinement of already formed interfaces. Once the main variant pattern has developed, PFNet advances the multichannel state with high local fidelity.

Rollout statistics in **Fig. 5b** show that the martensitic problem is more demanding than the CH case, yet remains tractable. The rollout error remains low over short horizons and increases gradually with rollout length, reaching the $10^{-1}$ range only at the longest tested horizon. This behavior reflects the greater complexity of multichannel AC-type phase-field evolution. Martensitic transformation involves stronger channel coupling, sharper anisotropy, and more sensitive synchronization among competing variants. A local phase error in one channel can therefore propagate to other channels through the coupled free-energy landscape. The observed error growth reflects the intrinsic difficulty of the problem rather than a failure of the one-step evolution operator.

Field-level comparisons in **Fig. 5c** support this interpretation. For the representative order parameter $\eta_1$, predictions at $\Delta t = 1$, 10, and 20 remain close to the reference solution, preserving the dominant twin-band orientation, spacing, and connectivity. At Δt=50, the main stripe skeleton is still retained, although local wall positions begin to decorrelate. Even at Δt=100, where fine-scale disorder becomes more visible, the prediction remains physically recognizable and does not collapse into an unstructured field. This result is notable because no martensite-specific redesign is introduced. The model remains effective after only increasing the channel dimension.

This benchmark further shows that PFNet is not tied to conserved scalar CH dynamics. The same framework can be extended directly from $C = 1$ to $C = 4$ without changing the core operator-learning formulation. The relevant object is therefore the microstructural state tensor rather than a scalar field alone. This observation is important for more general phase-field systems, where conserved compositions, non-conserved structural order parameters, multiple variants, multiple phases, or coupled field variables may coexist.

**Discussion**

We present PFNet as a physics-embedded surrogate framework for learning microstructural evolution operators. Rather than fitting short-time microstructure-to-microstructure correlations, PFNet approximates a conditional evolution operator over the microstructural state space. Its design integrates a variationally motivated formulation, entropy-based state conditioning, and a periodicity-preserving representation, enabling the model to advance phase-field microstructures with high local fidelity while maintaining stable long-horizon behavior.

For CH dynamics, PFNet provides a reliable approximation to coarsening trajectories over physically relevant rollout windows. Its performance remains robust across average composition, gradient-energy coefficient, coarsening stage, and morphology class. The model captures droplet-like structures, bicontinuous patterns, and inverted morphologies without being tied to a single morphology class or parameter setting. Errors are mainly confined to diffuse interfaces and topology-changing regions, while the large-scale coarsening pathway is preserved. These results demonstrate parameter-level and morphology-level generalization within conserved CH kinetics.

The martensitic benchmark further shows PFNet is not restricted to scalar conserved dynamics. By extending the state representation from one channel to four channels without changing the core backbone, PFNet remains effective for an AC-type, coupled multi-order-parameter system. This transition from conserved scalar CH coarsening to non-conserved multichannel martensitic transformation demonstrates generalization across distinct phase-field kinetics. The learned object is therefore not a problem-specific predictor of one equation, but an evolution operator acting on microstructural states.

These findings suggest that the transferability of PFNet arises from embedding physical structure into the surrogate at multiple levels. Periodic padding incorporates boundary consistency at the representation level. Entropy conditioning provides a trajectory-adaptive descriptor of instantaneous disorder. Parameter modulation enables the update rule to respond to changes in the free-energy landscape. These design principles link operator learning with thermodynamic descriptors, boundary consistency and variationally driven dynamics.

More broadly, PFNet illustrates how physical structure can be embedded at the representation, conditioning and operator levels to improve the reliability of learned simulators. Although demonstrated here using phase-field microstructure evolution, the underlying design principles are applicable to other dissipative systems governed by variational driving forces. This perspective connects phase-field surrogate modeling with the wider agenda of physics-informed machine learning: constructing predictive models that are data-efficient, physically interpretable and transferable across regimes.

Remaining challenges include improving long-horizon interfacial registration and cross-channel synchronization, particularly in stiff regimes and near topology-changing events. Future implementations may benefit from conservation-aware objectives, constraint-preserving updates and correction layers that enforce admissible state ranges during rollout. In this sense, PFNet provides a transferable step toward learning evolution operators for thermodynamically constrained dynamical systems.

## Funding statement

This work was financially supported by the National Natural Science Foundation of China (Grant Nos. 52401015, 12372152), Shanghai Artificial Intelligence Open Source Award Project, Guangdong Basic and Applied Basic Research Foundation (NO. 2024A1515012469), Shenzhen Science and Technology Program (JCYJ20250604124205007), and the Shandong Provincial Natural Science Foundation (ZR2023MA058).

## Method

### Phase-Field Simulation and Data Generation

The CH equation is adopted as the benchmark system to generate high-fidelity data. Its conserved dynamics enables systematic analysis of the physics-embedded architecture under a strict mass-conservation constraint.

The physical state of the Ti-Nb system is represented by the Nb concentration field $\phi(\boldsymbol{x}, t) \in [0, 1]$. Numerical integration is performed in an auxiliary variable $\psi(\boldsymbol{x}, t) = 2\phi(\boldsymbol{x}, t) - 1$. The microstructure evolution is then solved in terms of $\psi$ through

$$\frac{\partial \psi(\boldsymbol{x}, t)}{\partial t} = \nabla \left[ M \nabla \left( \frac{\partial f}{\partial \phi} - \kappa \nabla^2 \phi \right) \right] \tag{14}$$

where $M$ denotes the mobility and fixed to unity in this work. The chemical free energy density is

$$f(\psi) = RT \left[ \frac{1+\psi}{2} \ln \frac{1+\psi}{2} + \frac{1-\psi}{2} \ln \frac{1-\psi}{2} \right] + L_{\mathrm{TiNb}} \frac{(1+\psi)(1-\psi)}{4} \tag{15}$$

Here, $R$ is the gas constant, $T$ is the absolute temperature, and $L_{\mathrm{TiNb}} = 9.6\ \mathrm{kJ\ mol^{-1}}$ is the regular solution parameter describing the interaction between components Ti and Nb, and $\kappa$ is the gradient energy coefficient. The simulated solution is mapped back to the physical concentration field after numerical evolution according to $\phi(\boldsymbol{x}, t) = \left(1 + \psi(\boldsymbol{x}, t)\right)/2$

A uniformly spaced parameter grid was used to generate a diverse dataset. The initial phase fraction of niobium ($\phi_{\mathrm{Nb}}$) was set between 0.325 and 0.675 with an interval of 0.05, and $\kappa$ ranges from 4.5 to 11.5 with an interval of 1. For the test set, $\phi_{\mathrm{Nb}}$ ranged from 0.35 to 0.65 with an interval of 0.1 and $\kappa$ was selected from $\{5, 7, 9, 11\}$. Each simulated trajectory was converted into one-step input-target pairs $(\phi^t, \phi^{t+1})$, where $\phi^t \in \mathbb{R}^{1\times128\times128}$. The corresponding entropy $\mathcal{H}(\phi^t)$ was computed and used together with $\kappa$ to construct the conditional input.

**Training protocol**

PFNet is trained using a single-step prediction objective so that the model learns the local dynamics of the phase-field system. The training loss is defined as

$$\mathcal{L}_{\mathrm{step}} = \| G_\theta(\phi^t, \mathcal{H}(\phi^t), \kappa) - \phi^{t+1} \|_2^2 \tag{16}$$

Entropy is computed on-the-fly from each input sample, ensuring that the conditioning signal reflects the current microstructural state. Training was performed using the AdamW optimizer with an initial learning rate of $10^{-4}$ and weight decay of $10^{-4}$. A gradual warmup scheduler linearly increased the learning rate from the base value to $2 \times 10^{-4}$ over the first 10% of training epochs, followed by cosine annealing decay to zero over the remaining epochs. Gradient clipping with a maximum norm of one was applied to prevent exploding gradients. The batch size was set to 32, and training was conducted for 200 epochs with early stopping.

**Autoregressive rollout and evaluation metrics**

Although the model is trained on single-step predictions, its performance can be evaluated through multi-step autoregressive rollouts to assess long-term stability and accuracy. Given an initial condition, PFNet iteratively predicts future states, and the entropy is recomputed from the predicted state at each step. This autoregressive formulation mirrors the sequential nature of physical evolution and tests whether the model can maintain physical consistency over extended time horizons. The mean squared error (MSE) is defined as:

$$\mathrm{MSE} = \left\|\phi^{(T)} - \hat{\phi}^{(T)}\right\|^2 \tag{17}$$

where $\hat{\phi}^{(T)}$ is the prediction after $T$ autoregressive steps, and $\phi^{(T)}$ is the ground truth. The case $T$ =1 corresponds to one-step prediction, whereas larger $T$ values quantify long-horizon rollout accuracy.

Pointwise error maps were computed as

$$\mathrm{e}(\mathbf{x}) = \left|\phi(\mathbf{x}) - \hat{\phi}(\mathbf{x})\right|$$

and were used to identify spatial regions where prediction discrepancies were concentrated. For profile-level comparisons, concentration profiles were extracted across representative interfaces and compared between PFNet predictions and reference phase-field simulations.

**Martensitic-transformation benchmark**

The same framework was applied to a multichannel martensitic-transformation benchmark. In this case, the microstructural state was represented by four coupled non-conserved order-parameter fields, $\eta$=($\eta_1$, $\eta_2$, $\eta_3$, $\eta_4$) which describe multivariant martensitic domains. The evolution follows Allen-Cahn-type kinetics for non-conserved structural order parameters coupled through the underlying crystallographic and elastic constraints. Details of the martensitic phase-field model and simulation protocol follow previous studies [46–48].

The PFNet backbone, entropy-based conditioning, circular padding and autoregressive rollout strategy were retained. The only architectural change was the expansion of the input-output tensor from $B \times 1 \times H \times W$ to $B \times 4 \times H \times W$. The $\kappa$-conditioning branch was removed because variations in $\kappa$ were not considered in this benchmark. This setup tests whether PFNet learns a general microstructural evolution operator rather than a scalar-specific predictor for Cahn-Hilliard coarsening. For the martensitic benchmark, one-step prediction and autoregressive rollout were evaluated using the same mean squared error metric as in the Cahn-Hilliard case, with the error averaged over all order-parameter channels and spatial grid points.

## Supplementary Note 1: Detailed architecture and implementation of PFNet

PFNet is a U-Net-based surrogate model (~86M parameters) that maps a phase-field state to its next timestep in a single forward pass. The architecture adapts the denoising diffusion probabilistic model (DDPM) backbone by introducing two physics-informed modifications. All convolutions employ circular padding to enforce the periodic boundary conditions inherent to phase-field simulations, and the sinusoidal timestep embedding of DDPM is replaced by an entropy-based conditioning that encodes the microstructural disorder of the current state.

The training loss is evaluated in a linearly rescaled space to improve numerical conditioning. We use the AdamW optimizer with a learning rate of $10^{-4}$, weight decay of $10^{-4}$, and gradient clipping at a maximum norm of 1.0. The learning rate follows a linear warmup over 20 epochs to twice the base value, then decays to zero via cosine annealing over the remaining 180 epochs. All models are trained for 200 epochs with a batch size of 32.

**Table S1**. Data normalization for different task

| Task | Physical range | Model scale |
|---|---|---|
| CH | [0, 1] | $\phi \times 12 + 8$ |
| AC | [-1, 1] | $\eta \times 6 + 6$ |

**U-net backbone**

The encoder consists of four levels. Each level contains two residual blocks followed by a strided circular-padded convolution for Downsampling (except the final level). Channel widths are [128, 256, 384, 512], determined by multipliers [1, 2, 3, 4] applied to the base width ch = 128. Self-attention is applied only at the 384-channel level (Level 2), where the spatial resolution ($32 \times 32$ for $128 \times 128$ input) balances computational cost and receptive field. The bottleneck comprises two residual blocks at 512 channels, the first includes self-attention and the second does not.

The decoder mirrors the encoder with four levels, each containing three residual blocks (one more than the encoder to accommodate concatenated skip connections) followed by nearest-neighbor upsampling with a circular-padded convolution. Self-attention in the decoder is applied at Level 2 in all three residual blocks, maintaining symmetry with the encoder. The complete architecture contains 22 residual blocks (8 encoder, 2 bottleneck, and 12 decoder) and 11 skip connections.

**Circular padding**

All $3 \times 3$ convolutions use circular padding to enforce periodic boundary conditions. This applies to the head convolution (1), both convolutions in each residual block ($22 \times 2 = 44$), downsampling convolutions (3), upsampling convolutions (3), and the tail convolution (1), totaling 52 circular padding operations per forward pass. The $1 \times 1$ convolutions including skip connections and attention projections do not require padding.

**Downsampling and upsampling**

Downsampling uses a circular-padded $3 \times 3$ convolution with stride 2. Upsampling uses nearest-neighbor interpolation ($\times 2$) followed by a circular-padded $3 \times 3$ convolution.

**Entropy conditioning**

The Shannon image entropy is computed from a 256-bin histogram per channel and averaged, then clamped to [1.5, 3.0]. The scalar is embedded via 32 Gaussian radial basis functions with centers uniformly spaced in [1.5, 3.0] and bandwidth (3.0 - 1.5)/31, followed by a linear projection to 128 dimensions. The resulting vector is broadcast spatially and added to the head features.

## $\kappa$ conditioning via FiLM

The gradient energy coefficient $\kappa$ in [3.0, 15.0] is embedded identically to entropy (32 RBF centers in [3.0, 15.0]) but projected to 512 dimensions. This embedding is passed to all 22 residual blocks, where per-block linear layers project it to FiLM parameters.

## Residual block with FiLM

Each residual block contains two convolution stages with a single FiLM modulation inserted between them:

$$h_1 = \mathrm{Conv}(\mathrm{Swish}(\mathrm{GroupNorm}(h_{\mathrm{in}})))$$

$$[\gamma, \beta] = \mathrm{split}(\mathrm{Swish}(\mathrm{Linear}(\mathrm{temb})))$$

$$h_2 = \mathrm{Conv}(\mathrm{Dropout}(\mathrm{Swish}(\mathrm{GroupNorm}(h_1) \odot (1 + \gamma) + \beta)))$$

$$h_{\mathrm{out}} = \mathrm{Attn}(h_2 + \mathrm{Shortcut}(h_{\mathrm{in}}))$$

where Shortcut is a $1 \times 1$ convolution when channel dimensions change, otherwise identity. The FiLM parameters $\gamma, \beta \in \mathbb{R}^{B \times C_{\mathrm{out}}}$ are derived from the conditioning embedding $\mathrm{temb} \in \mathbb{R}^{B \times 512}$ via a per-block linear projection. When $\mathrm{temb} = 0$ (non-$\kappa$ tasks), FiLM degenerates to plain GroupNorm. Dropout ($p = 0.15$) is applied only in the second stage. Self-attention, when active, uses GroupNorm pre-normalization with scaled dot-product attention and a residual connection.

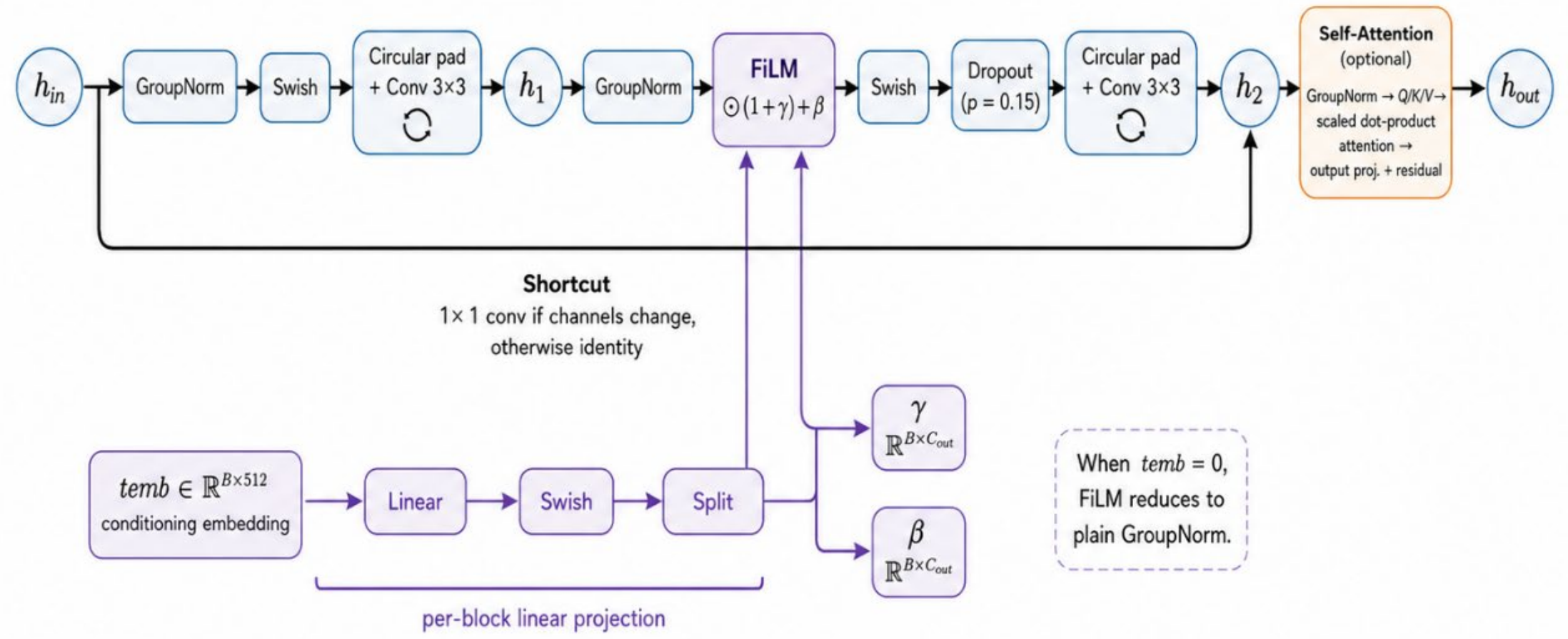

**Fig. S1** Residual block architecture with FiLM-based parameter conditioning. This design enables the model to dynamically modulate its internal representations based on the gradient-energy coefficient $\kappa$, facilitating parameter-aware surrogate modeling across different interfacial energy landscapes without task-specific retraining.

## Weight initialization

All convolutional and linear layers use Xavier uniform initialization with zero bias. Output projection layers (the second convolution in each residual block, the attention output projection, and the tail convolution) use Xavier uniform with gain $10^{-5}$ to ensure near-zero initial output.

**Supplementary Note 2: Effect of $\kappa$ and $\phi_0$ on structure evolution**

The gradient-energy coefficient $\kappa$ in the Cahn-Hilliard equation directly controls the interfacial width and energy penalty $\mathcal{F}[\phi] = \int_{\Omega}\left(f(\phi) + \frac{\kappa}{2}|\nabla\phi|^2\right)d\mathbf{x}$, where $f(\phi)$ is the chemical free energy density and the second term represents the interfacial energy contribution. The characteristic interfacial width scales as $\delta \sim \sqrt{\frac{\kappa}{f''(\phi_{eq})}}$ where $f''(\phi_{eq})$ is the second derivative of the free energy at the equilibrium composition. A smaller $\kappa$ narrows the diffuse interface width and steepens local concentration profiles. These localized variations make the subsequent evolution more difficult to resolve, especially during early-stage amplification and interface sharpening.

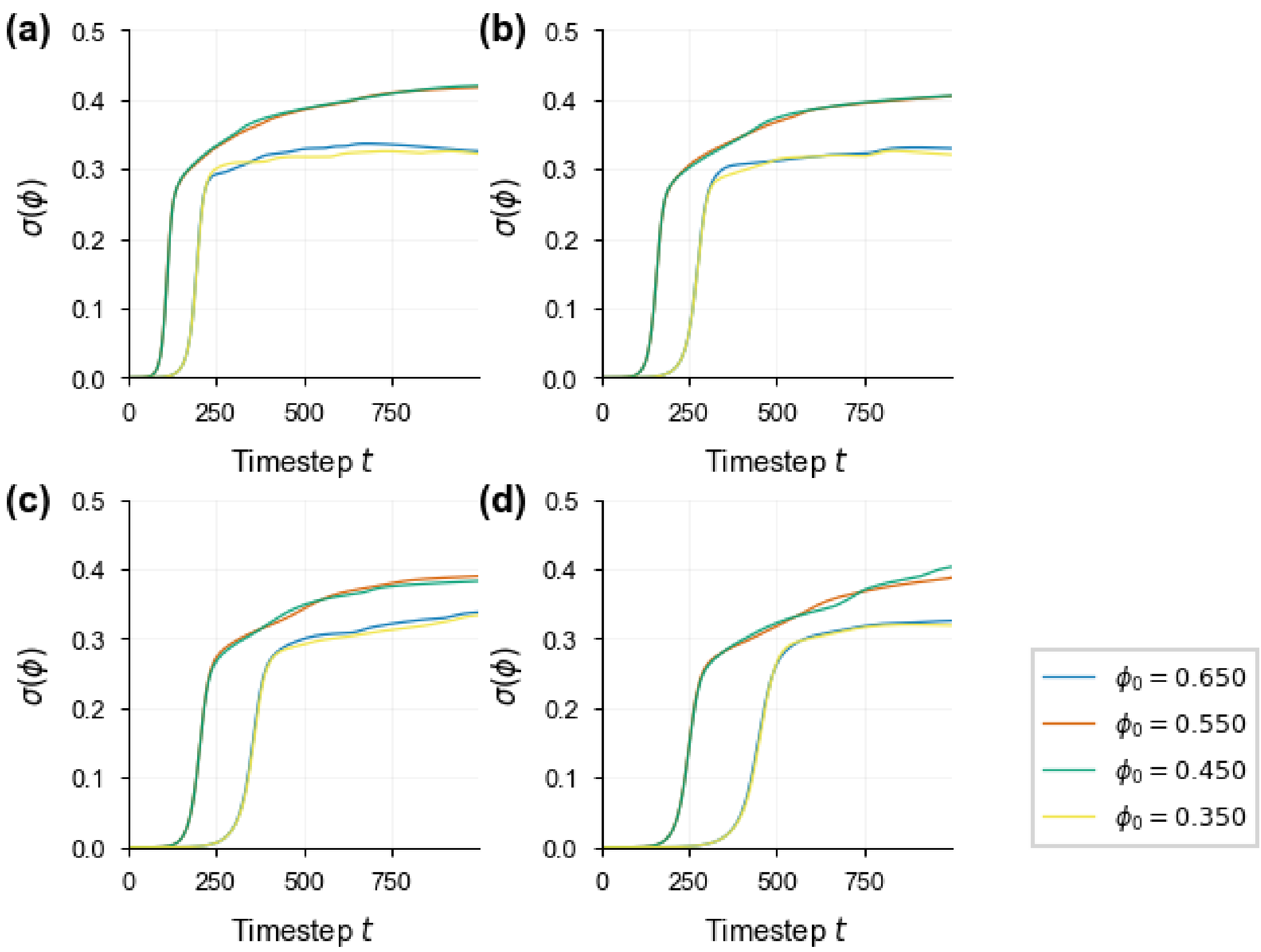

Fig. S2. Temporal evolution of the spatial standard deviation $\sigma(\varphi)$ for different initial compositions $\varphi_0$ and gradient-energy coefficients $\kappa$. Panels (a–d) correspond to $\kappa$ = 5, 7, 9, and 11, respectively. Smaller $\kappa$ leads to earlier growth of concentration fluctuations, indicating faster spinodal decomposition and sharper interface formation.

**Fig. S2** summarizes the temporal evolution of the spatial standard deviation $\sigma(\phi)$

$$\sigma(\phi) = \sqrt{\frac{1}{N^2}\sum_{i,j}\left(\phi_{ij} - \frac{1}{N^2}\sum_{i,j}\phi_{ij}\right)^2}$$

for different $\kappa$ and initial average compositions $\phi_0$. For all compositions, decreasing $\kappa$ accelerates the development of concentration fluctuations. The onset time, defined here as the first timestep at which $\sigma(\phi)$ exceeds 0.05, increases systematically with $\kappa$. For example, for $\phi_0 = 0.550$ and 0.450, the onset occurs at approximately t ≈ 95 for $\kappa = 5$, but is delayed to t ≈ 220 for $\kappa = 11$. This confirms that smaller $\kappa$ promotes faster amplification of spinodal modes and sharper early interfacial development.

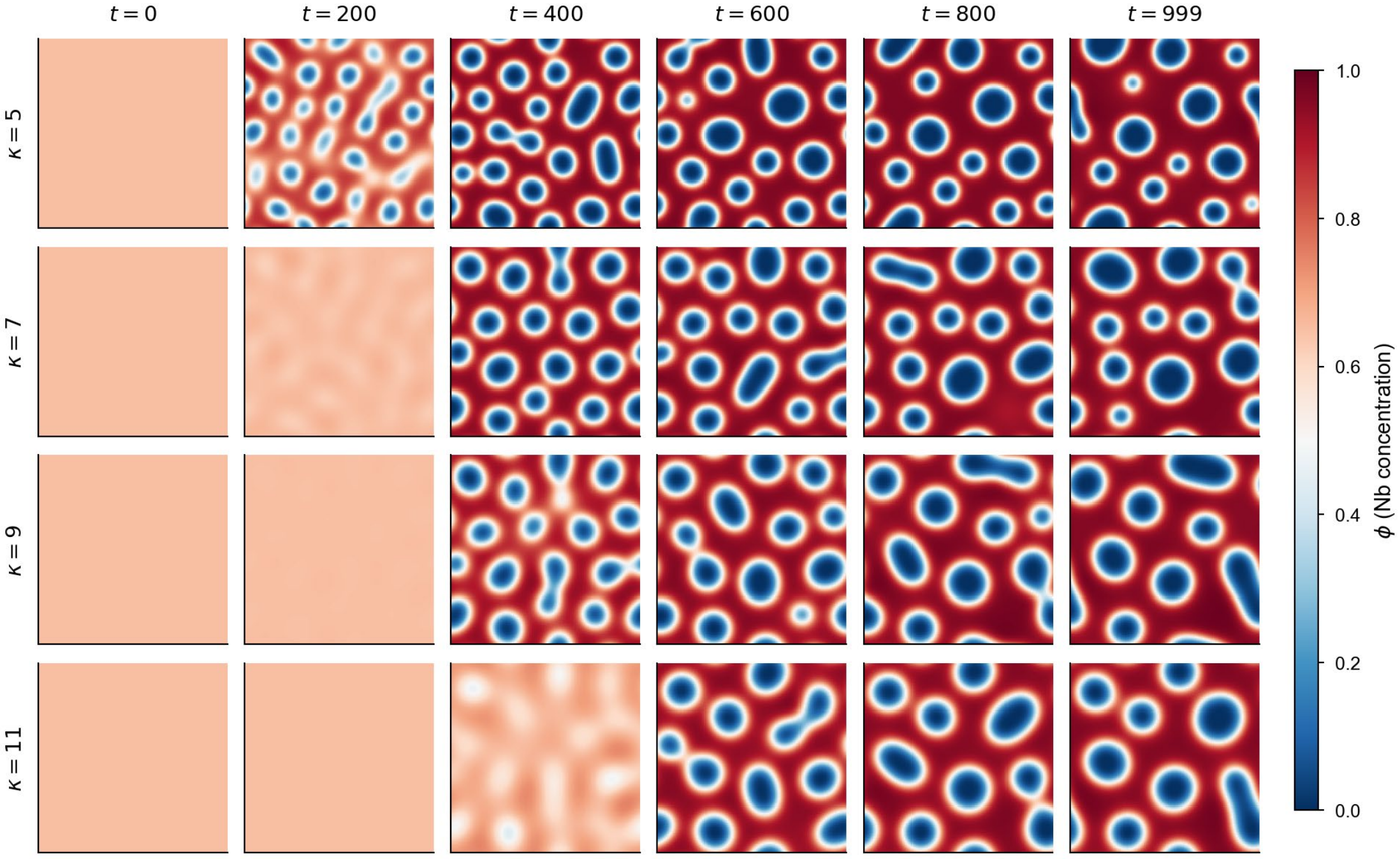

**Fig. S3**. Microstructure evolution for $\phi_0 = 0.650$. Rows correspond to $\kappa = 5$, 7, 9, and 11; columns correspond to $t = 0$, 200, 400, 600, 800, and 1000. The system evolves toward a high-$\phi$ matrix containing low-$\phi$ droplets. Smaller $\kappa$ produces sharper interfaces and earlier phase separation.

The microstructure snapshots in **Figs. S3-S6** provide direct visual evidence of the $\kappa$ dependence. At smaller $\kappa$, phase boundaries are visibly sharper and concentration transitions occur over fewer grid spacings. This behavior is especially clear at $\kappa = 5$, where high- and low-concentration domains rapidly separate into well-defined regions by $t = 200$ and 400. By contrast, larger $\kappa$ values delay the emergence of strong contrast and maintain broader diffuse interfaces for longer times. For a fixed initial composition, increasing $\kappa$ slows the transition from weak concentration fluctuations to fully phase-separated structures. In the $\kappa$ = 11 cases, the microstructure remains close to the initial homogeneous state for a longer period, and clear domain contrast appears later than in the corresponding $\kappa$ = 5 and $\kappa$ = 7 cases.

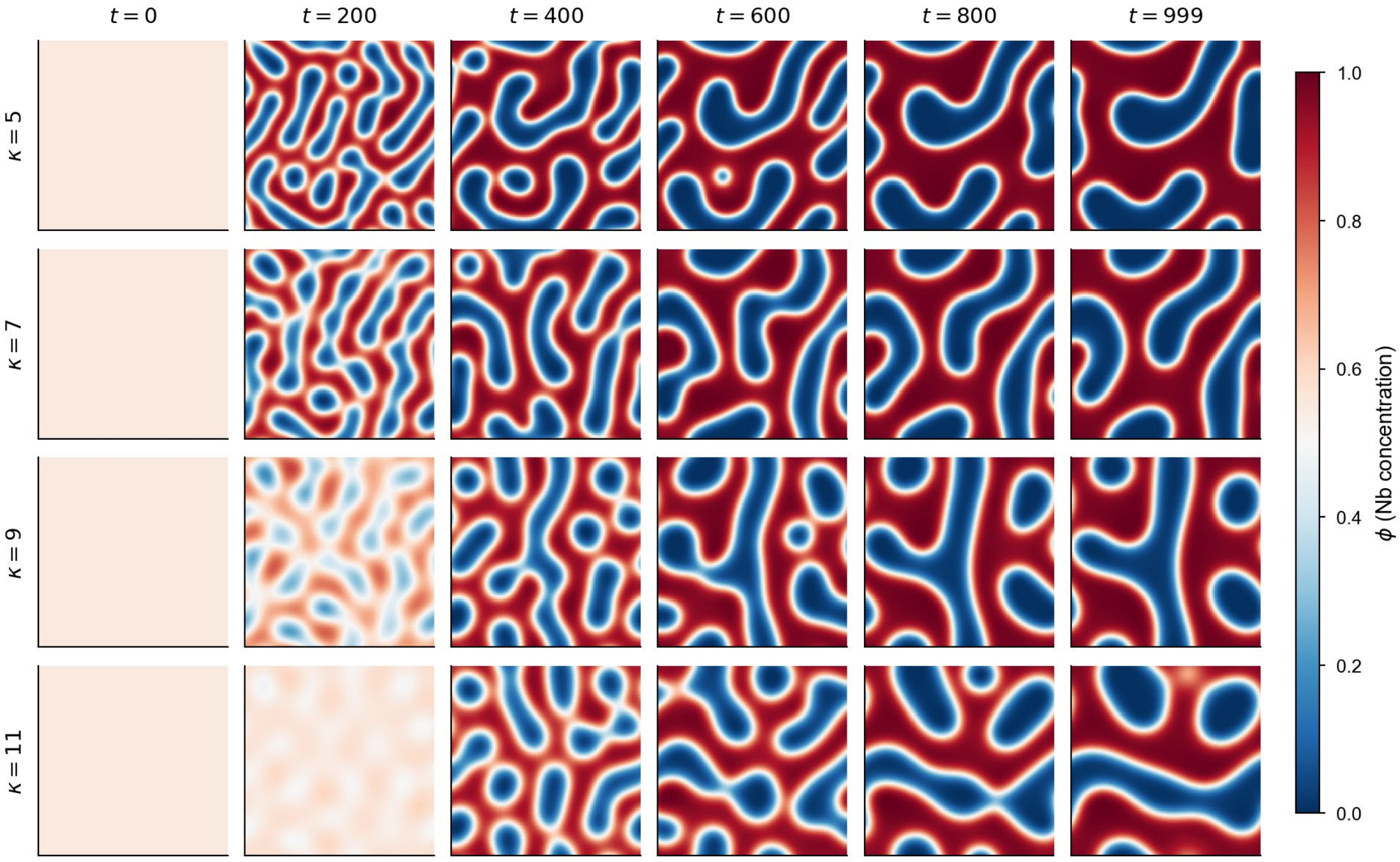

**Fig. S4**. Microstructure evolution for $\phi_0$ = 0.550. Near-equiatomic composition produces a bicontinuous, labyrinthine morphology. Compared with off-critical compositions, this morphology involves stronger nonlinear rearrangement and coarsening of interconnected domains.

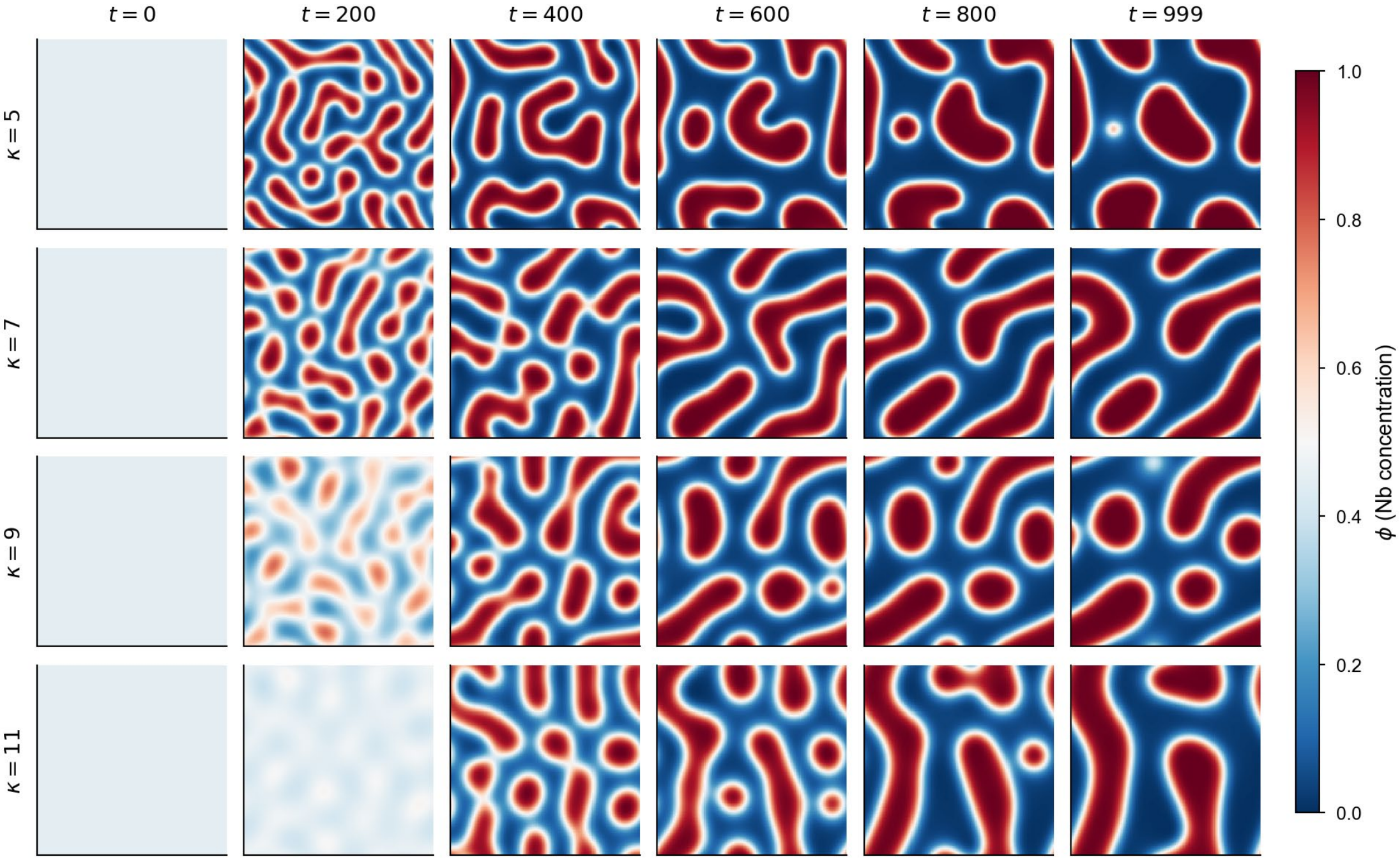

**Fig. S5**. Microstructure evolution for $\phi_0$ = 0.450. The morphology is approximately complementary to the $\phi_0$ = 0.550 case, with interconnected high- and low-concentration domains.

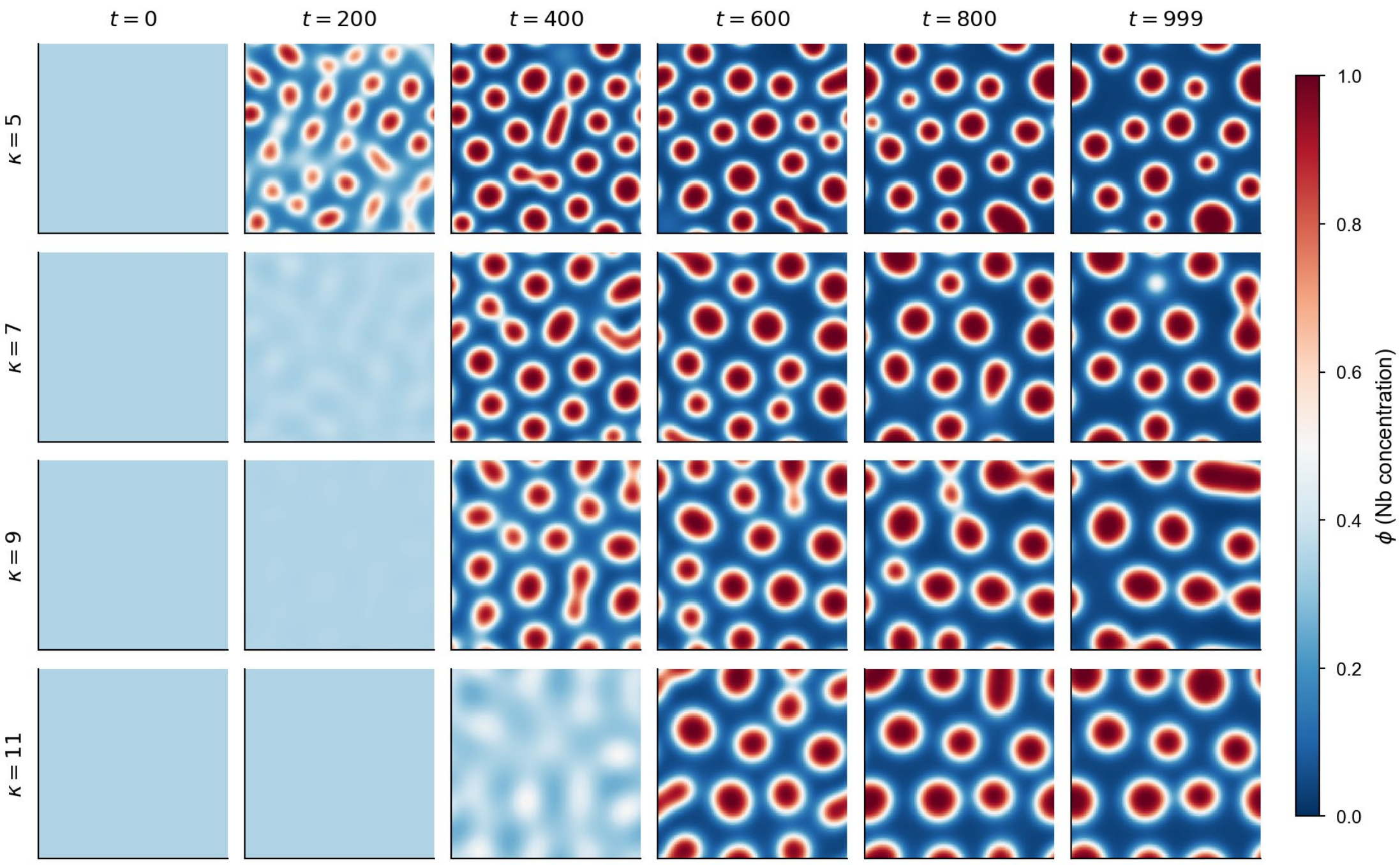

**Fig. S6**. Microstructure evolution for $\phi_0 = 0.350$. The system evolves toward low-$\phi$ matrix regions containing high-$\phi$ droplets. As in the $\phi_0 = 0.650$ case, the off-critical composition favors droplet-like phase separation rather than a fully bicontinuous network.